\documentclass[journal=ancac3,manuscript=article]{achemso}

\usepackage{chemformula} 
\usepackage[T1]{fontenc} 
\usepackage{siunitx}
\usepackage{gensymb}
\usepackage{anyfontsize}
\usepackage{lineno}
\usepackage{newtxtext,newtxmath,amsmath}

\author{Peter Kepič}
\affiliation[CEITEC]
{Brno University of Technology, Central European Institute of Technology, Purky\v{n}ova 123, \\ 612 00, Brno, Czech Republic}

\author{Petra Kalousková}
\affiliation[UFI]
{Brno University of Technology, Faculty of Mechanical Engineering, Technická 2, \\ 616 69, Brno, Czech Republic}

\author{Tomáš Šikola}
\affiliation[CEITEC]
{Brno University of Technology, Central European Institute of Technology, Purky\v{n}ova 123, \\ 612 00, Brno, Czech Republic}
\alsoaffiliation[UFI]
{Brno University of Technology, Faculty of Mechanical Engineering, Technická 2,  \\ 616 69, Brno, Czech Republic}

\author{Filip Ligmajer}
\email{filip.ligmajer@ceitec.vutbr.cz}
\affiliation[CEITEC]
{Brno University of Technology, Central European Institute of Technology, Purky\v{n}ova 123, \\ 612 00, Brno, Czech Republic}
\alsoaffiliation[UFI]
{Brno University of Technology, Faculty of Mechanical Engineering, Technická 2, \\ 616 69, Brno, Czech Republic}

\title[Article Title]{Controlled dewetting and phase transition hysteresis of VO$_2$ nanostructures}


\begin{document}

\begin{abstract}

As artificial intelligence continues to grow, so does the need for more efficient ways to process data. Besides moving from electronic to photonic circuits, a promising approach is to integrate phase-change materials. Vanadium dioxide (VO$_2$) exhibits an ultrafast, near-room-temperature phase transition, characterized by hysteresis and large optical modulation---making it a promising candidate for short-term memories and for mimicking neural behavior in brain-like computing systems. While the hysteresis behavior of VO$_2$ has been well studied in thin films and nanostructures, practical control and device integration have been limited only to thin films. Here, we demonstrate control over the phase transitions of VO$_2$ nanocylinders via lithographic patterning, controlled crystallization, and controlled dewetting. Because nanostructures are easier to address and consume less power than films, the ability to fabricate them with tailored geometry and hysteresis properties directly on integrated platforms is a key step toward scalable, energy-efficient memory and neuromorphic photonic devices.

\end{abstract}

\section{Introduction}

The continuous advancement of artificial intelligence and quantum computation is driving increasing demand for energy-efficient, fast-switching, and scalable memory and processing technologies. The most promising direction involves transitioning from traditional electronic integrated circuits to photonic ones. Out of all material platforms for storing and processing information in photonic circuits, phase-change materials excel due to their significant modulation of optical properties and low energy consumption during the phase transition~\cite{Lian2022}. Among them, vanadium dioxide (VO$_2$) stands out as one of the fastest switching ones, with around \SI{100}{fs} insulator-metal transition (IMT)~\cite{Lopez2004, Ocallahan2015} that can be achieved electrically, optically or thermally already around \SI{68}{\celsius}~\cite{Morin1959, Kepic2021}. Such a low transition temperature facilitates fabrication of memory devices made of microscale VO$_2$ films that require only picojoules of energy to switch~\cite{Jung2022, Parra2025}. Besides the ultrafast and low-energy transition, the VO$_2$ phase transition exhibits hysteresis behavior, meaning the metal-insulator transition (MIT) occurs at lower temperatures than IMT, around \SI{60}{\celsius} for bulk VO$_2$. While insufficient for long-term memories, as energy must be constantly supplied to store the information in the metallic phase, it is a good candidate for short-term memories and brain-like artificial neurons~\cite{Schofield2023, Lee2024}. In such devices, the information can be written by a switching pulse and read by low-power pulses that follow within the \SI{e1}{}--\SI{e2}{ns} window after~\cite{Rini2005, Ryckman2013}. To reduce the switching or retention energy and prolong the retention time, the MIT and overall hysteresis characteristics can be tailored by adjusting stoichiometry, strain, volume, defects, and doping~\cite{Andrews2019}. The effect of those properties on hysteresis has been extensively studied and exploited in thin VO$_2$ films, where one can, for example, decrease the transition temperature by tungsten dopants~\cite{Kaufman2024, White2025} or shift the transition temperature~\cite{Breckenfeld2017} and increase the transition steepness~\cite{White2021} via strain from a lattice-matched substrate. Hysteresis characteristics have also been adjusted by controlling defects in lithographically patterned polycrystalline nanostructures~\cite{Appavoo2012}, which exhibit hysteresis similar to that of bulk materials. Interestingly, single-crystal VO$_2$ nanoparticles (NPs), with minimal crystallographic defects, exhibit MITs as low as \SI{30}{\celsius}~\cite{Lopez2002, Donev2009, Clarke2018, Nishikawa2023}. This effect brings the entire memory platform closer to room temperature with the apparent benefit of energy efficiency. Benefiting from the recently reported possibility of controlling the transition temperatures by the size of such NPs~\cite{Kepic2025}, a memory made of them could host several levels of information in an area that is more than one order of magnitude smaller than in films. Nevertheless, this promising hysteresis-broadening size effect has been limited to NPs of random size and distribution after dewetting the VO$_2$ film, i.e., after thermally inducing a continuous film to break into isolated droplets via surface energy minimization.

\begin{figure*}
    \centering
    \includegraphics{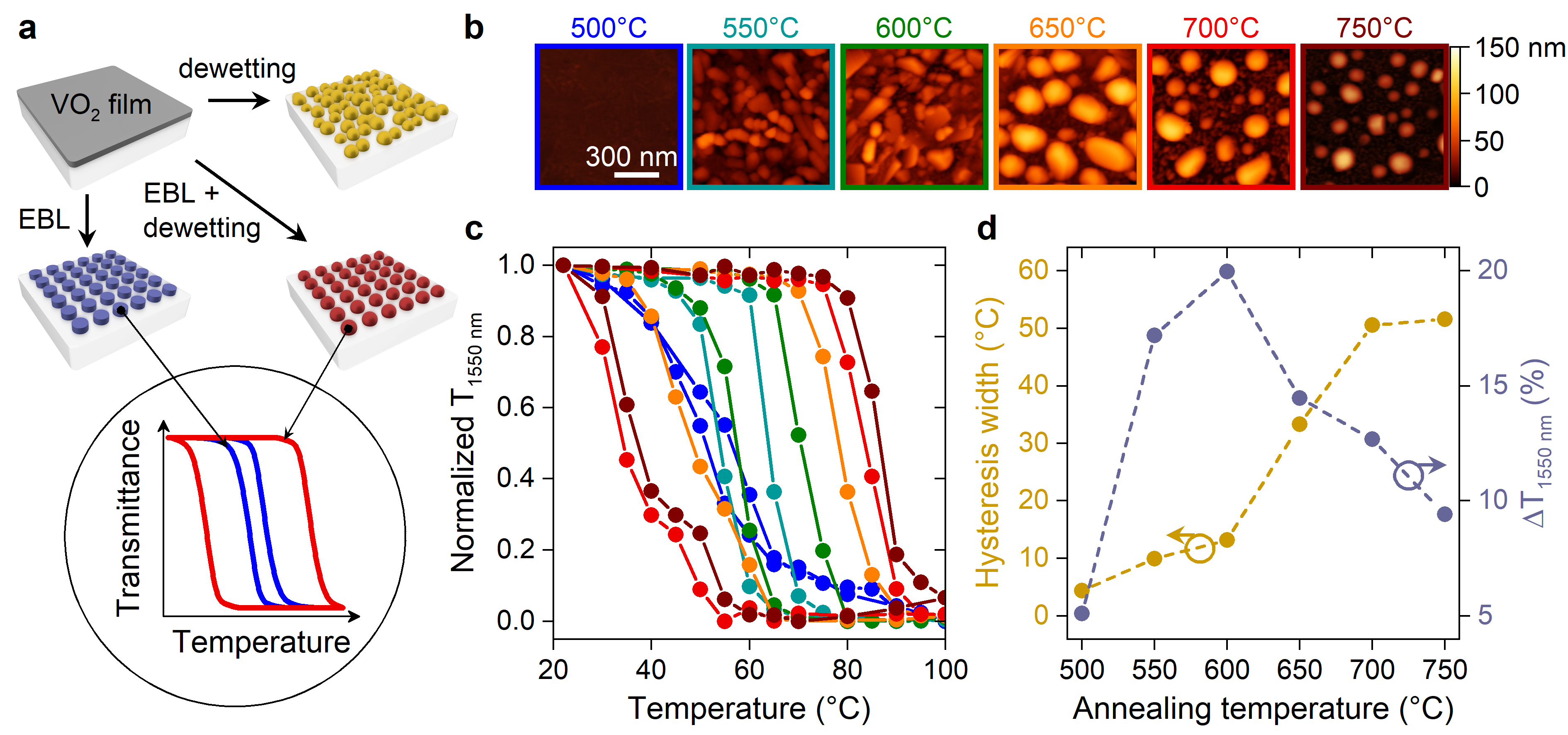}
    \caption{Dewetting of VO$_2$ films. a) Schematic of the phase transition hysteresis broadening by dewetting bare VO$_2$ films and lithographically patterned VO$_2$ nanocylinders. b) AFM topography of \SI{50}{nm} VO$_2$ films annealed for \SI{10}{min} at selected temperatures. c) The thermal hysteresis in normalized transmittance at \SI{1550}{nm} wavelength of films shown in (b). The lines' colors correspond to the micrographs' border colors. d) Hysteresis width (left axis) and $\Delta \text{T}_\text{1550 nm}$ (right axis) of the films in (b) as functions of the respective annealing temperatures.}
    \label{Figure1}
\end{figure*}

Here, we conduct a systematic study of crystallization and dewetting of lithographically patterned VO$_2$ nanocylinders and the effect of size and annealing on the hysteresis characteristics (Figure~\ref{Figure1}a). As a reference, we first examine unpatterned VO$_2$ films to understand baseline hysteresis behavior under annealing. We then turn to lithographically patterned VO$_2$ nanocylinders with varying diameters after annealing, studying how annealing-induced dewetting alters their morphology and phase transition behavior. Finally, we correlate nanostructure size and annealing conditions with transition temperatures, hysteresis width, and optical modulation. This way, we create a library from which nanocylinders with desired hysteresis parameters can be chosen and fabricated. Such a library is essential for the design and fabrication of energy-efficient VO$_2$ nanostructures for multilevel memory and neuromorphic photonic devices.

\section{Results and discussion}

\textbf{Phase transition hysteresis of VO$_2$ films.} Depositing VO$_2$ with the desired switching properties is challenging due to a strong dependence of these properties on stoichiometry and crystalline quality~\cite{Griffiths1974, Shvets2019}. These material attributes are primarily governed by an interplay between oxygen pressure and temperature during deposition~\cite{Rampelberg2015, Shibuya2015} and by parameters of ex-situ annealing~\cite{Marvel2015}. To demonstrate the effect of annealing temperature, we took six samples of amorphous VO$_2$ films (\SI{30}{nm} thick; see Methods in Supporting Information) and exposed them to six different temperatures in a vacuum furnace for \SI{10}{min} under \SI{15}{sccm} oxygen flow. As the films were deposited onto amorphous fused silica substrates and annealed ex situ (post-deposition), they exhibited a polycrystalline character (Figure~S1). Figure~\ref{Figure1}b shows that by increasing the annealing temperature above \SI{500}{\celsius}, the VO$_2$ grains become coarser (from \SI{1}{nm} to \SI{14}{nm} RMS roughness). Moreover, above \SI{600}{\celsius}, the films dewet into individual VO$_2$ NPs, which then further aggregate. Finally, at \SI{750}{\celsius}, these NPs begin to wet (spread) and seemingly shrink due to oxidation toward the more volatile V$_2$O$_5$ stoichiometry, which exhibits a lower contact angle and forms a “skirt” around the nanoparticle~\cite{Suh2004, White2025}. Note that a similar behavior was also reported recently for continuous films annealed at prolonged times~\cite{White2025}. The process of thin-film dewetting into separate NPs is strongly dependent on the initial film thickness, with thicker films requiring higher annealing temperatures and times, or even rendering dewetting practically impossible~\cite{White2025}. The increasing annealing temperature leads not only to these morphological changes, but also to an increase in IMT temperature and to a decrease in MIT temperature, effectively broadening the hysteresis (Figure~\ref{Figure1}c). Note that transition temperatures and the hysteresis width are defined in Methods in Supporting Information. The film annealed at the lowest temperature of \SI{500}{\celsius} exhibits very narrow and gradual hysteresis with only \SI{4}{\celsius} width (see Figure~\ref{Figure1}d) and IMT shifted below \SI{60}{\celsius}. This gradual transition is associated with a large number of small crystal grains, consistent with prior studies, which found that smaller grain sizes led to more gradual phase transitions~\cite{Alcaide2023, Marvel2013}. The shifted IMT, on the other hand, is a result of the reduction of stoichiometry towards V$_2$O$_3$ (see X-ray photoelectron spectra in Figure~S1). For films annealed at \SI{550}{\celsius} and \SI{600}{\celsius}, the hysteresis got steeper and broader (\SI{10}{\celsius} and \SI{13}{\celsius}, respectively), reflecting the larger grains with fewer defects. The hysteresis width of films annealed at \SI{700}{\celsius} and \SI{750}{\celsius} jumps to \SI{50}{\celsius}, as a result of the film dewetting into separated NPs. The film annealed at \SI{650}{\celsius} contained a mix of NPs and clusters of grainy films (see Figure~S1), and thus exhibited an intermediate hysteresis width of \SI{33}{\celsius}. To investigate how these annealing effects reflect in the photonic properties of VO$_2$, we also looked at the transmittance difference between the insulator and metallic phases within the telecom wavelength range ($\Delta \text{T}_\text{1550 nm}$). Figure~\ref{Figure1}d shows that small grains and imperfect stoichiometry at \SI{500}{\celsius} annealing temperature also translate to only \SI{5}{\percent} optical modulation. With improved crystallinity at larger annealing temperatures, $\Delta \text{T}_\text{1550 nm}$ significantly increases up to \SI{20}{\percent}. However, it decreases again as the film starts to dewet, dropping to \SI{9}{\percent}. As the VO$_2$ transforms into smaller isolated NPs, the overall transmission increases and $\Delta \text{T}_\text{1550 nm}$ decreases, since NPs cover less surface and in the metallic phase support a localized surface plasmon resonance (LSPR) that shifts out of the \SI{1550}{nm} wavelength as size decreases (discussed in more detail later in the text). Thus, although hysteresis continues to broaden with smaller NPs, the optical contrast diminishes.

This dependence of hysteresis widths and $\Delta \text{T}_\text{1550 nm}$ confirms two physical effects related to regimes of grainy films and separated NPs~\cite{White2025}: In grainy but continuous films (500--\SI{650}{\celsius}), large grains with fewer defects result in broader hysteresis, as there are fewer oxygen vacancies at grain boundaries to initiate the nucleation of one phase in another~\cite{Appavoo2012}. After the film dewets into separated NPs, its hysteresis continues to broaden as the grain boundaries and defects almost disappear~\cite{Appavoo2012, Lopez2002, Nishikawa2023}. In the regime of separated NPs (between 650--\SI{750}{\celsius}), as the NPs get smaller (see Figure~S1), the hysteresis further broadens, as the amount of nucleation centers decreases with the decreasing size~\cite{Lopez2002, Nishikawa2023, Clarke2018}. The path towards hysteresis broadening is thus two-fold: smaller isolated NPs or larger grains within the polycrystalline film. However, in the context of optical modulation, these two strategies lead to opposite results: While the grain growth within VO$_2$ films at medium annealing temperatures increases $\Delta \text{T}_\text{1550 nm}$, the subsequent dewetting and shrinking of separated NPs at larger annealing temperatures decrease it. As the NPs get smaller, the overall transmittance in both VO$_2$ phases increases due to their smaller surface coverage. But because the transmittance increase in the metallic phase is larger than that in the insulating phase (due to suppression of LSPR~\cite{Appavoo2012, Kepic2021, Kepic2025}), $\Delta \text{T}_\text{1550 nm}$ decreases with smaller NPs. Understanding these regimes and controlling widths and $\Delta \text{T}_\text{1550 nm}$ by adjusting the annealing temperature is important when either directly incorporating VO$_2$ thin films into integrated or far-field nanophotonic devices, or creating VO$_2$ nanostructures.

\begin{figure}
    \centering
    \includegraphics{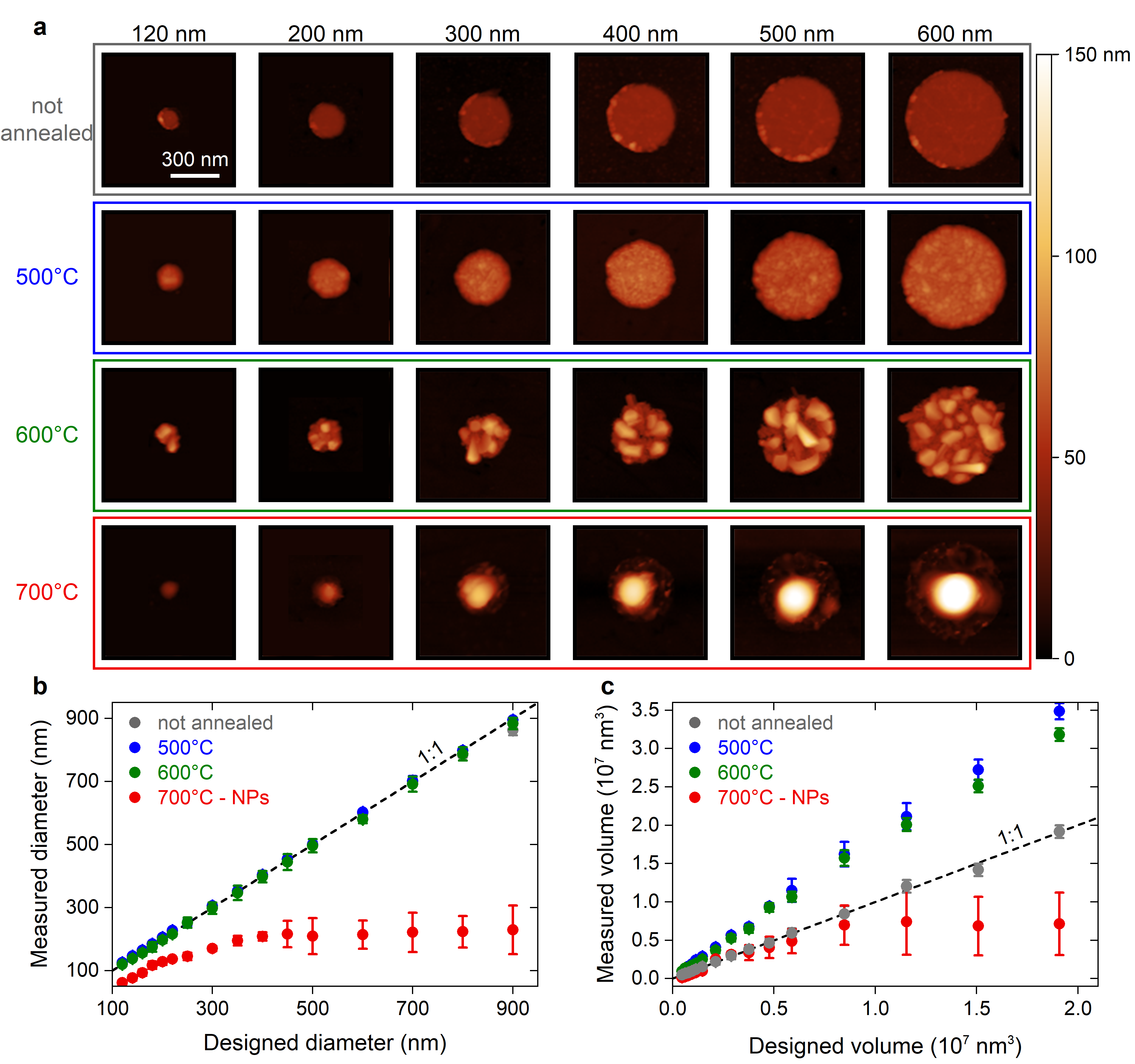}
    \caption{Morphology of VO$_2$ nanocylinders before and after annealing in the oxygen atmosphere. a) AFM topography of VO$_2$ nanocylinders with the listed designed diameters before and after annealing at the listed temperatures for \SI{10}{min} under \SI{15}{sccm} oxygen flow. Note that although the nanocylinders with \SI{120}{nm} and \SI{200}{nm} diameters are arranged in an array with a 1.5$\times$D spacing (implying multiple structures within the 800$\times$\SI{800}{nm} view-field), only the central nanocylinder is shown for visual clarity. b) Measured diameter as a function of the designed diameter. c) Measured volume as a function of the designed volume (nanocylinders of a \SI{30}{nm} height). The dashed lines in (b) and (c) represent a 1:1 ratio between the designed and measured quantities. The error bars represent a standard deviation of $\geq$10 nanocylinders.}
    \label{Figure2}
\end{figure}

\textbf{Morphology of VO$_2$ nanocylinders.} To have sufficient control over VO$_2$ hysteresis at nanoscale volumes and to utilize the size-dependent phase transitions of dewetted NPs, it is important first to understand morphological changes of lithographically-patterned nanostructures after annealing. Note that fabricating nanostructures by etching them from already annealed VO$_2$ films was not explored because their granular structure (see Figure~\ref{Figure1}b) severely limits the achievable lateral resolution. We fabricated six samples, each containing arrays of amorphous VO$_2$ nanocylinders with diameters ranging from 120 to \SI{900}{nm}. The center-to-center spacing was \SI{1.5}{\times} the cylinder diameter, and each array covered a 20$\times$\SI{20}{\micro m^2} area (see Methods in Supporting Information). We then annealed each sample in a vacuum furnace for \SI{10}{min} under \SI{15}{sccm} oxygen flow and a different temperature (one sample per temperature). In the first row of Figure~\ref{Figure2}a, we can see the as-fabricated (i.e., not annealed) VO$_2$ nanocylinders with a very smooth surface (ca. \SI{1}{nm} roughness) and occasional rougher edges caused by fabrication imperfections. Their measured diameters are as designed (represented by grey data points in Figure~\ref{Figure2}b), and the measured volume also corresponds to the designed height of \SI{30}{nm} (Figure~\ref{Figure2}c). When annealed at \SI{500}{\celsius}, the shape and diameter of nanocylinders remain unchanged, but the surface roughness increases to ca. \SI{3}{nm}, while the height increases to \SI{57}{nm} (represented by the increased volume in Figure~\ref{Figure2}c). VO$_2$ thus grows almost \SI{2}{\times} in the vertical direction during annealing at \SI{500}{\celsius}. This growth can be surprising, since crystalline VO$_2$ is \SI{1.6}{\times} denser than amorphous VO$_2$~\cite{Wen2013, Wu2023}, and the structure should shrink upon crystallization, not expand. We thus ascribe the observed growth to additional oxidation of any imperfect stoichiometries and to the formation of pores between grains, as confirmed by SEM cross-sections that show a more porous columnar structure after annealing, consistent with volumetric expansion due to oxidation (Figure~S1). The similar vertical growth is also confirmed for nanocylinders annealed at \SI{550}{\celsius} (see Figure~S2). At \SI{600}{\celsius}, even larger grains are formed within the nanocylinders (roughness increases to ca. \SI{25}{nm}), but the diameter remains within \SI{5}{\percent} of the as-designed value (Figure~S2), and the lithographic definition holds up. The height of the cylindrical envelope then reaches \SI{50}{nm}, which is still \SI{1.7}{\times} larger than before annealing. The situation changes at \SI{700}{\celsius}, where VO$_2$ nanocylinders suddenly dewet and form NPs. Around them, a very thin (5$\pm$5)nm non-uniform polycrystalline residue remains in the original footprint of the cylinder (essentially a faint 'shadow' of the former nanocylinder, see Figure~\ref{Figure2}a). The diameters of dewetted NPs are approximately \SI{60}{\percent} of the initial nanocylinders, up to the diameter of \SI{400}{nm}. Beyond this threshold, their average diameter asymptotically approaches \SI{220}{nm}, while their variance increases significantly due to dewetting of these larger nanocylinders into more than one NP (see below). All these dewetted NPs are roughly hemispherical or, more precisely, prolate spherical caps with their height equal to \SI{60}{\percent} of the diameter. Their volumes follow a similar trend to their diameters: Up until the designed diameter of \SI{500}{nm} (\SI{0.6e7}{nm^3}), their volumes are slightly reduced to \SI{77}{\percent} of the volumes before annealing. For larger diameters, their average volume no longer depends on the designed volume and saturates around \SI{0.7e7}{nm^3}, with the increased variance (Figure~\ref{Figure2}c). Similar behavior, but shifted to smaller measured diameters, can be observed for NPs obtained after dewetting the nanocylinders at even higher temperature of \SI{750}{\celsius} (see Figure~S2). These results indicate that the morphological changes of nanocylinders after annealing are analogous to those of the films: Until the annealing temperature is high enough to cause dewetting into NPs, the diameters of nanocylinders after annealing are unaffected, while their heights/volumes grow significantly. Such growth should be considered when fabricating nanophotonic devices that utilize plasmonic or Mie resonances in nanostructures of specific sizes \cite{Kepic2025}. When the annealing temperature is high enough to cause dewetting, nanocylinders evolve into hemispherical NPs, whose sizes follow the trends shown in Figure~\ref{Figure2}, and can be directly linked to trends in hysteresis properties, as shown below. 

\begin{figure}
    \centering
    \includegraphics{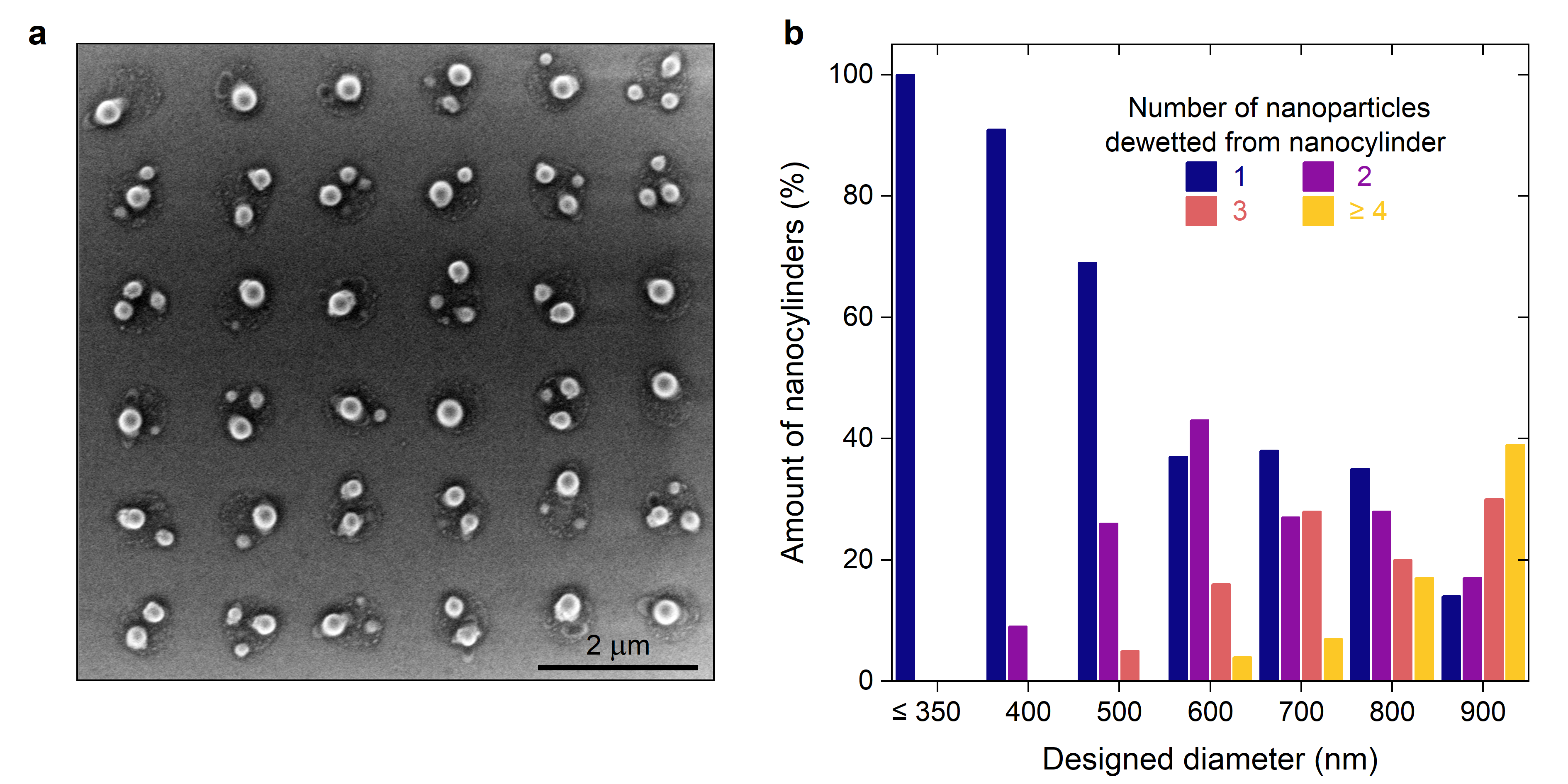}
    \caption{Amount of NPs dewetted from one VO$_2$ nanocylinder. a) Large field-of-view SEM micrograph of VO$_2$ nanocylinders with the \SI{700}{nm} diameter annealed at \SI{700}{\celsius} for \SI{10}{min}. b) Percentage of nanocylinders that dewetted into 1, 2, 3, or $\geq$4 NPs after annealing at \SI{700}{\celsius} as a function of the designed diameter.}
    \label{Figure3}
\end{figure}

To have full control over the dewetting of nanocylinders into specific NPs, it is important to know how many NPs are formed by the dewetting of nanocylinders of specific diameters. As already shown in Figure~\ref{Figure2}b, the average final diameter of dewetted NPs after annealing at $\geq$\SI{700}{\celsius}  does not exceed \SI{220}{nm}, regardless of the nanocylinder's diameter. The large standard deviation in diameters also indicates considerable variability in NP diameters under these conditions. We quantified this variability by processing SEM micrographs taken at larger fields-of-view (Figure~\ref{Figure3}a) and extracted the statistics of the number of NPs into which nanocylinders dewetted after annealing at \SI{700}{\celsius}. In Figure~\ref{Figure3}b, we can see that all nanocylinders of diameters up to \SI{350}{nm} dewetted into single NPs with diameters shown in Figure~\ref{Figure2}b. For the \SI{400}{nm} diameter, \SI{92}{\percent} of nanocylinders dewetted into one NP, while \SI{8}{\percent} dewetted into two NPs. Above \SI{500}{nm}, patterned nanocylinders behaved similarly to unpatterned films, breaking apart into a random number of NPs. While even \SI{900}{nm} nanocylinders can be dewetted into single NPs with a diameter of approximately \SI{400}{nm}, the probability of such an event is only \SI{15}{\percent}. In practical terms, \textasciitilde\SI{220}{nm} is thus the upper limit of the single NP diameter that can be reliably obtained from a single starting feature. Together with the morphological changes and diameters of NPs described in the previous paragraph, we have gathered all the information necessary to controllably fabricate individual VO$_2$ NPs of a given diameter at a given position.

\begin{figure}
    \centering
    \includegraphics{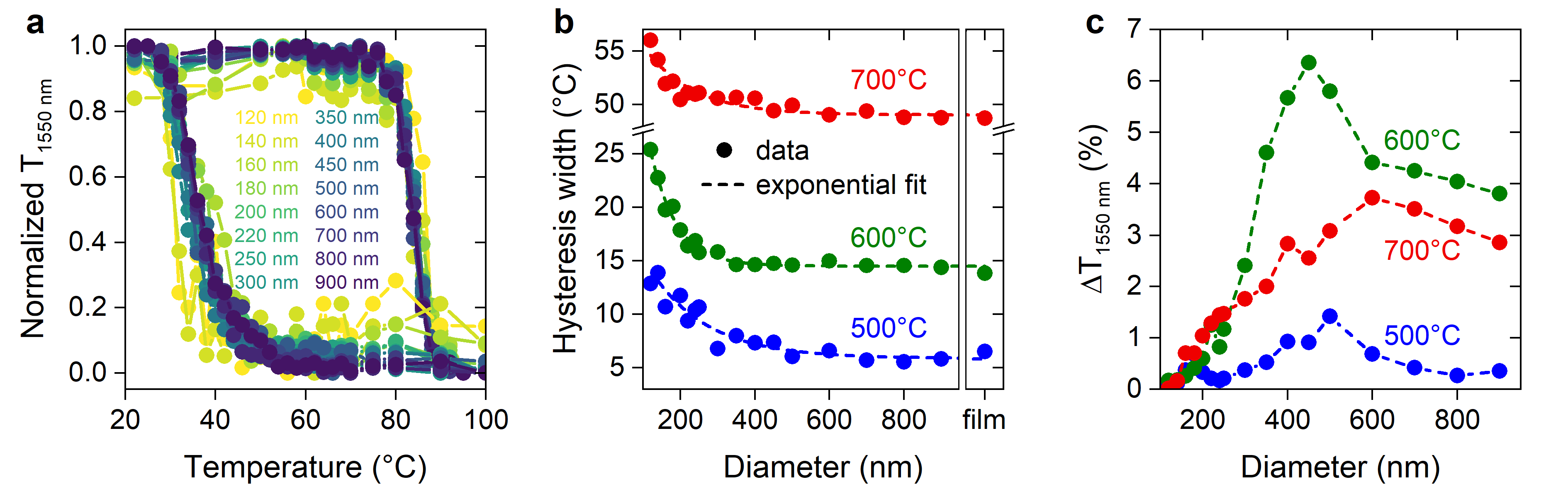}
    \caption{The phase transition hysteresis of VO$_2$ nanocylinders. a) The temperature hysteresis of normalized $\text{T}_\text{1550 nm}$ of the listed VO$_2$ nanocylinders annealed at \SI{700}{\celsius} for \SI{10}{min}. b) Hysteresis width and c) $\Delta \text{T}_\text{1550 nm}$ of VO$_2$ nanocylinders annealed at \SI{500}{\celsius}, \SI{600}{\celsius}, and \SI{700}{\celsius} as functions of the designed nanocylinder diameter. Note that absolute $\Delta \text{T}$ values for nanocylinders appear lower than for films in Figure~\ref{Figure1} due to different measurement setups and fill-factors, so comparisons should be just qualitative.}
    \label{Figure4}
\end{figure}

\textbf{Phase transition hysteresis of VO$_2$ nanocylinders.} To quantify the hysteresis properties of the annealed VO$_2$ nanocylinder arrays, we analyzed their optical properties by means of infrared camera images at \SI{1550}{nm} wavelength during a series of heating and cooling cycles (see Methods and Video~1 in Supporting Information). Figure~\ref{Figure4}a shows the normalized transmittance of the nanocylinder arrays annealed at \SI{700}{\celsius} (see Figure~S3 for other temperatures). We can observe the characteristic hysteresis, \SI{48}{\celsius} wide for the largest NPs, which is broadening towards \SI{56}{\celsius} as the diameter decreases. Increasing the IMT temperature and decreasing the MIT temperature are the two factors behind the broadening (see detailed analysis in Figure~S4). The hysteresis width is further quantified in Figure~\ref{Figure4}b, where it is plotted as a function of the designed nanocylinder diameter. Specifically, the hysteresis width decreases from \SI{56}{\celsius} at \SI{120}{nm} diameter to \SI{50}{\celsius} at \SI{400}{nm} diameter, and ultimately to \SI{48}{\celsius}, when nanocylinders dewet into multiple NPs. A similar exponential decrease of the hysteresis widths towards larger diameters can be observed for nanocylinders annealed at \SI{500}{\celsius} (from \SI{14}{\celsius} to \SI{6}{\celsius}), \SI{600}{\celsius} (from \SI{25}{\celsius} to \SI{14}{\celsius}), and other temperatures (see additional datasets in Figure~S4). This exponential size-dependence of hysteresis might seem counter-intuitive at first (since the physical NP size in Figure~\ref{Figure2}b increased roughly linearly with design), but it is consistent with known phase-transition nucleation theory~\cite{Lopez2002, Appavoo2012}: In essence, smaller or more pristine VO$_2$ crystals require more supercooling/overheating to transition (hysteresis is broad) because they lack internal defects that can seed the new phase. Larger or polycrystalline structures contain more such nucleation sites (e.g., oxygen vacancies at grain boundaries), so they switch at temperatures closer to equilibrium, yielding a narrower hysteresis. The hysteresis width of non-dewetted nanocylinders annealed at \SI{500}{\celsius} and \SI{600}{\celsius} is more than \SI{3}{\times} narrower than that of NPs due to a significantly larger amount of defects at grain boundaries~\cite{Lopez2002, Appavoo2012}. Nevertheless, nanocylinders still follow the same size-dependence that can be fitted by a decreasing exponential function based on the theory of heterogeneous phase transitions~\cite{Lopez2002, Nishikawa2023} (see the dashed lines in Figure~\ref{Figure4}b). From an application perspective, the ability to tune the hysteresis width via geometry (diameter) is more important than the absolute hysteresis width itself. Although the tuning range of \SI{10}{\celsius} hysteresis width reported here for nanocylinder arrays annealed at \SI{600}{\celsius} is smaller than that of individual single-crystal NPs we reported previously~\cite{Kepic2025}, it confirms the proposed idea of encoding information into the size of nanostructures and demonstrates its practical feasibility. With this controlled dewetting approach, we can cover nearly the whole space of transition temperatures between 49--\SI{95}{\celsius} (IMT) and 30--\SI{63}{\celsius} (MIT) by combining the two degrees of freedom: diameter and annealing temperature (see Figure~S4). Nevertheless, the large-scale annealing demonstrated here crystallizes all nanostructures at once. To truly address the hysteresis width of specific individual nanostructures, one would have to utilize techniques such as local selective crystallization by pulsed laser~\cite{Li2025} or spatial light modulation. 

In addition to the tunability of hysteresis width, we also investigated the tunability of infrared transmission at telecom wavelengths. Figure~\ref{Figure4}c shows $\Delta \text{T}_\text{1550 nm}$ as a function of the designed diameter for the same samples as in Figure~\ref{Figure4}b. We can see that overall $\Delta \text{T}_\text{1550 nm}$ of the nanocylinders annealed at \SI{500}{\celsius} is mostly below \SI{1}{\percent}, then it increases even above \SI{6}{\percent} at \SI{600}{\celsius} and ultimately drops below \SI{4}{\percent} when nanocylinders are dewetted into NPs at $\geq$\SI{700}{\celsius} (see also Figure~S4). Therefore, for the fixed diameter of nanocylinders, their $\Delta \text{T}_\text{1550 nm}$ follows the annealing temperature trend of the films in Figure~\ref{Figure1}d. In other words, the broader hysteresis comes at the cost of smaller modulation (for a given nanocylinder diameter). To maintain the same transmission modulation levels, a larger footprint, caused by a greater number of NPs, will be required. Note that although $\Delta \text{T}_\text{1550 nm}$ of nanocylinders seems to be more than \SI{3}{\times} smaller than that of the films, these data cannot be directly compared: The former was obtained from IR microscopy images utilizing focused light, while the latter was obtained from IR spectroscopy utilizing collimated laser and large-spot averaging. Besides the overall transmission modulation by means of annealing temperatures discussed above, $\Delta \text{T}_\text{1550 nm}$ in Figure~\ref{Figure4}c exhibits pronounced modulation as a function of diameter. Conventionally, one would expect bigger VO$_2$ elements to block more light in the metallic state (hence larger $\Delta \text{T}_\text{1550 nm}$). However, metallic VO$_2$ nanocylinders support LSPR that redshifts with the increasing size~\cite{Appavoo2012, Kepic2021, Kepic2025}. For nanocylinders above 500--\SI{600}{nm}, this resonance is shifted sufficiently that at \SI{1550}{nm} the NPs are less lossy and transmission in the metallic phase actually rises for large diameters. This causes $\Delta \text{T}_\text{1550 nm}$ to peak at specific diameters and then diminish for larger sizes (as confirmed by the results of numerical simulations shown in Figure~S5). This decrease in modulation magnitude indicates another strength of nanostructuring VO$_2$: By choosing an optimal size of nanocylinders, one can harness LSPR to achieve a relatively high per-particle $\Delta \text{T}$ while still maintaining a reasonably broad hysteresis, thereby needing fewer particles overall than a film patch would, significantly decreasing the footprint of a potential integrated device. Regardless of the specific goal --- be it maximizing optical modulation, selecting transition temperatures, or tailoring hysteresis widths --- our study provides a library of VO$_2$ nanostructures that can act as potential building blocks for integrated multilevel memories and neuromorphic photonic devices.

\section{Conclusions}

In summary, we established a systematic relationship between annealing-driven morphological evolution and phase-transition characteristics in lithographically defined VO$_2$ nanostructures, providing both mechanistic insight and a practical route to device-relevant tuning. Using unpatterned films as a reference, we observed grain coarsening with increasing annealing temperature up to \SI{600}{\celsius}, the onset of dewetting above \SI{600}{\celsius}, and ultimately size reduction due to oxidation at \SI{750}{\celsius}. These morphological changes were accompanied by pronounced phase transition hysteresis broadening from \SI{4}{\celsius} to \SI{52}{\celsius} as the morphology transitioned from a continuous, grainy film to isolated, dewetted NPs. Extending this approach to amorphous VO$_2$ nanocylinder arrays, we observed analogous regimes and a reproducible >\SI{1.7}{\times} increase in volume during crystallization, occurring predominantly along the vertical direction. We further identified a deterministic dewetting window at \SI{700}{\celsius}: nanocylinders with diameters up to \SI{400}{nm} predominantly collapse into a single NP, whose diameter saturates near \SI{220}{nm}, whereas larger nanocylinders dewet stochastically into multiple particles. Optical hysteresis measurements at \SI{1550}{nm} reveal a robust, size-dependent tuning of hysteresis width that decays exponentially with the increasing nanocylinder diameter (with convergence to film-like behavior above \SI{400}{nm}), while the transmittance modulation follows the same annealing trends as films and can be significantly enhanced by LSPR for 400--\SI{600}{nm} diameters. The broad hysteresis thus comes at the cost of lower optical contrast, and designers must decide between maximizing hysteresis width (memory-state stability) and optical contrast. Our results show these are opposing effects for a given VO$_2$ volume. Collectively, these results demonstrate scalable control over position, crystallization state, dewetting outcome, and phase-transition metrics, yielding a library of VO$_2$ nanostructures. These nanostructures can then be selected based on specific figure-of-merit metrics (transition temperatures, hysteresis widths, modulation depths), which are crucial for future compact, energy-efficient, tunable, and neuromorphic photonic architectures.

\begin{acknowledgement}

This work is supported by the Grant Agency of the Czech Republic (project No. 25-18336M) and by the Ministry of Education, Youth and Sports of the Czech Republic (project No. LUABA24069). We acknowledge the CzechNanoLab Research Infrastructure (ID 90251), funded by MEYS CR, for the financial support of the measurements and sample fabrication. We also thank our colleague Dr. Petr Bouchal for his generous help with the optical measurement setup.

\end{acknowledgement}

\begin{suppinfo}

\begin{itemize}
  \item Supporting Information: Methods; The additional morphology and stoichiometry of VO$_2$ films; The additional morphology, normalized $\text{T}_\text{1550 nm}$, phase transition temperatures, hysteresis widths and $\Delta \text{T}_\text{1550 nm}$ of VO$_2$ nanocylinders; Transmittance spectra and $\Delta \text{T}_\text{1550 nm}$ simulations.
  \item Video1: Sequences of IR camera images capturing the phase transition hysteresis of VO$_2$ NP arrays annealed at \SI{700}{\celsius}.
\end{itemize}

\end{suppinfo}

\bibliography{bibliography}

\end{document}